\documentclass[11pt,a4paper]{article}

\usepackage[T1]{fontenc}
\usepackage{mathptmx}
\usepackage{booktabs}
\usepackage{amsmath,float,longtable}
\usepackage{array}
\usepackage{caption}
\usepackage[dvipsnames]{xcolor}
\usepackage[a4paper,margin=1in]{geometry}
\usepackage{microtype}
\usepackage[round,authoryear]{natbib}
\usepackage{setspace}
\usepackage{titlesec}
\usepackage{url}
\usepackage{hyperref}

\definecolor{brandburgundy}{RGB}{128,0,32}
\hypersetup{
  colorlinks=true,
  linkcolor=brandburgundy,
  citecolor=brandburgundy,
  urlcolor=brandburgundy,
  pdftitle={Does Crypto Sentiment Extremity Widen Estimated Spreads? Evidence Depends on the Specification},
  pdfauthor={Murad Farzulla}
}
\titleformat{\section}{\large\bfseries\color{brandburgundy}}{\thesection}{0.5em}{}
\titlespacing*{\section}{0pt}{1.5ex}{0.7ex}
\newcolumntype{L}[1]{>{\raggedright\arraybackslash}p{#1}}

\begin{document}

\begin{center}
{\footnotesize\color{gray}\textsc{Dissensus Working Paper DAI-2510 | September 2026}}\\[1em]
{\LARGE\bfseries Does Crypto Sentiment Extremity Widen Estimated Spreads?}\\[0.35em]
{\Large\itshape Evidence Depends on the Specification}\\[1.1em]
{\large Murad Farzulla}\\[0.3em]
{\small Dissensus, London, UK}\\[0.2em]
{\footnotesize \href{mailto:murad@dissensus.ai}{murad@dissensus.ai}\quad
ORCID: \href{https://orcid.org/0009-0002-7164-8704}{0009-0002-7164-8704}}
\end{center}

\begin{abstract}
We examine whether extreme values of the Crypto Fear \& Greed Index are associated with a daily
high--low spread estimate for Bitcoin. The sample contains 2,896 BTC/USDT observations from
February 2018 to January 2026. We find an unconditional extreme-minus-neutral gap of 61.99 basis points.
After close-to-close realised-volatility-quintile demeaning it is 24.79 basis points, although none
of the five separate quintile contrasts survives Holm correction. With quadratic realised-volatility
and strictly lagged momentum controls, the HAC estimate is 11.81 basis points (95\% CI
$[-2.31,25.93]$, $p=.101$). A fixed non-parametric stratification gives 20.44 basis points
($p=.0195$ under circular shifts), while separate models for a zero-floored estimate's incidence and
positive magnitude are imprecise. The results therefore show only a descriptive,
specification-dependent association. We conclude that they do not establish a stable or causal
liquidity premium.


\end{abstract}

\section{Introduction}

Extreme investor sentiment has been associated with cryptocurrency herding
\citep{jia2022extreme}, while non-linear sentiment relationships have been reported for Bitcoin
volatility \citep{bourghelle2022emotions} and cross-coin price synchronicity
\citep{wang2024ushaped}. These outcomes make extremity a plausible empirical state variable, but
they do not establish that extremity changes transaction costs. Theory provides a possible lens:
informational asymmetry can generate a positive bid--ask spread \citep{glosten1985bid}. Our daily
data neither identify informed trading nor observe dealers' quotes, so that mechanism remains
motivation rather than an estimated channel.

This letter asks a narrower question: is an extreme-minus-neutral difference present in a
high--low spread estimate after accounting for volatility, recent returns, serial dependence, and
the estimator's mass at zero? The unresolved empirical problem is whether a positive raw difference
persists across defensible adjustments. The question arose from auditing an earlier, stronger manuscript.
That audit found that its momentum paragraph analysed Parkinson volatility rather than spreads and
that reported ``Cohen's $d$'' values used an unweighted average of group variances. We repair
both defects and show the resulting inferential disagreement. The result is positive
under every displayed specification, but its magnitude contracts and its statistical conclusion
depends on the adjustment. Our contribution is this contingency, not a new structural premium.

\section{Data and design}

We merge public Binance BTC/USDT daily OHLCV observations with Alternative.me's public aggregate
Crypto Fear \& Greed score by UTC calendar date. The resulting panel runs from 21 February 2018 to
29 January 2026 ($N=2{,}896$). Alternative.me describes the index as combining market and
attention inputs, including volatility, maximum drawdown, momentum/volume, social media, surveys,
Bitcoin dominance and Google Trends \citep{alternative2026}. The current documentation marks
surveys as paused. It publishes neither component histories nor a fully reconstructible, versioned
formula. We therefore treat the score as an aggregate market-state/sentiment proxy, not a pure
sentiment measure.

Our primary regime coding defines extreme fear as scores $\leq25$, fear as 26--45, neutral as
46--55, greed as 56--75, and extreme greed as $>75$. These are author-defined bands. Extreme fear
and extreme greed are pooled and compared with neutral, giving 888 extreme and 457 neutral days.
The API-provided class labels treat 47--54 as neutral in the cached panel; we report this alternative
classification as a sensitivity analysis.

The outcome is the Corwin--Schultz spread estimate in basis points. It uses adjacent daily high--low
ranges to separate volatility from a spread component \citep{corwin2012simple}. This is an estimated
daily-data spread, not an observed quoted or effective spread and not an adverse-selection measure.
The implementation floors negative estimates at zero. There are 935 such observations (32.3\%),
so we analyse both spread levels and, as a censoring sensitivity, whether the estimate is positive
and its log magnitude conditional on being positive.

We first report the unconditional mean gap and the earlier pooled comparison after demeaning within
five quintiles of 20-day annualised close-to-close realised volatility. We also report all five
separate quintile comparisons with Holm correction. The repair then uses the 2,876 observations
available after constructing momentum as the sum of the previous 20 log returns; the outcome day's
return is excluded. A quadratic OLS model includes standardised realised volatility, its square,
lagged momentum and its square. Moderate fear and greed form the common reference, and the estimand
is the extreme-indicator coefficient minus the neutral-indicator coefficient. Inference uses
Newey--West covariance with 20 ordered-observation lags. The two-part sensitivity uses a binomial
logit and OLS for the log positive estimate under the same controls and covariance design.

As a deliberately different adjustment, we fix a $5\times3$ realised-volatility-quintile by
momentum-tercile grid. We take the harmonic-sample-size-weighted average of within-cell
extreme-minus-neutral gaps and compare it with all 2,875 non-zero circular shifts of the full regime
sequence. This preserves the observed outcome order and regime run lengths but relies on
phase-exchangeability and does not cure omitted-state bias. Every analysis in this reconstruction is
post hoc and exploratory. The repaired design was fixed before its output files were persisted, but
it was not preregistered or independent of earlier inspection of the panel. All tests are two-sided.
The fixed grid contains as few as five neutral observations in a custom-class cell (three under the
provider classes), which limits cell-level stability.

OpenAI Codex assisted with drafting, code auditing and analysis-script development. The release
checks reconstruct the processed panel from cached exchange and index inputs, compare the custom
HAC calculation with an independent library implementation, and regenerate the reported outputs.

\section{Results}

Table~\ref{tab:ladder} presents the specification ladder. The raw gap is 61.99 basis points
($d=.374$ using the sample-size-weighted pooled standard deviation). Demeaning within
close-to-close volatility quintiles reduces the pooled gap to 24.79 basis points ($d=.154$). Its
IID pooled test is small, but the five separate gaps all have confidence intervals crossing zero
and none survives Holm correction. The pooled row is therefore exploratory aggregation, not five
replications.

\begin{table}[t]
\centering
\caption{Extreme-minus-neutral specification ladder}
\label{tab:ladder}
\footnotesize
\begin{tabular}{L{4.0cm}rL{3.25cm}L{2.7cm}r}
\toprule
Specification & $N$ & Effect & 95\% interval & $p$ \\
\midrule
Unconditional mean gap & 2,896 & +61.99 bps ($d=0.374$) & [46.60, 77.89] & $2.70\times 10^{-14}$ \\
RV-quintile demeaning & 2,896 & +24.79 bps ($d=0.154$) & -- & 0.0015 \\
Quadratic RV + momentum, HAC & 2,876 & +11.81 bps & [-2.31, 25.93] & 0.101 \\
Fixed 5$\times$3 stratification & 2,876 & +20.44 bps & -- & 0.0195 \\
Positive-estimate incidence, HAC & 2,876 & OR 1.045 & [0.806, 1.356] & 0.739 \\
Positive-estimate magnitude, HAC & 1,950 & +11.3\% & [-4.8\%, 30.2\%] & 0.180
\\
\bottomrule
\end{tabular}
\caption*{\footnotesize \textit{Notes:} The first row uses an IID Welch test and a seeded
percentile-bootstrap interval. The second uses an IID Welch test on pooled within-quintile
residuals; its five component tests give 0/5 Holm rejections. ``Quadratic'' controls for linear and
squared close-to-close realised volatility and strictly lagged 20-day momentum, with Newey--West
inference. The stratified row reports the complete circular-shift reference test and has no
confidence interval. The last two rows decompose the Corwin--Schultz floor; OR denotes odds ratio.
All analyses use the custom 46--55 neutral band and are exploratory.}
\end{table}

In the repaired parametric model, the adjusted gap is 11.81 basis points (95\% CI
$[-2.31,25.93]$, $p=.101$). It is roughly one-fifth of the raw difference and is not
distinguishable from zero at conventional levels. The fixed stratified analysis yields 20.44 basis
points with a circular-shift $p=.0195$. These are different estimands and different maintained
assumptions; the latter does not turn the former null into corroboration.

The floor decomposition is also weak. The point estimate for positive-estimate incidence is close to one, with a wide interval (OR 1.045, 95\% CI $[0.806,1.356]$, $p=.739$). Among
positive estimates, the adjusted difference is 11.3\% (95\% CI $[-4.8\%,30.2\%]$, $p=.180$).
Using API-provided classes changes the parametric spread result to 11.36 basis points
($p=.119$) and the stratified result to 20.12 basis points ($p=.0282$); the substantive tension is
unchanged.

Additional sensitivity reinforces the dependence on method. When the outcome is demeaned along a
range-based volatility axis, the full-sample IID $p$-value moves from .0063 with five bins to .061
with 20 bins and .108 with 50 bins. This axis shares high--low information with the spread
estimator, so it is diagnostic rather than a cleaner control. The full bin grid, individual
quintiles, model coefficients and classification sensitivity appear in the supplement.

\section{Interpretation and limitations}

The defensible finding is a specification-sensitive residual association. It is not evidence that
market makers observe this index, withdraw liquidity, or respond causally to sentiment. The index
itself embeds volatility and momentum/volume inputs. Our close-to-close controls proxy those market
states but cannot reconstruct or remove the provider's unpublished components. Contemporaneous
date matching also supports association, not prediction.

Proxy validity is the second boundary. This reconstruction has no independently verified quoted-
or effective-spread validation. Earlier comparisons labeled as such used trade-based proxies:
one Binance measure was a trimmed deviation from a rolling trade-price median, and files labeled
Bybit order books contained trades without bid/ask quotes. Those comparisons are withdrawn as
validation evidence. The analysis therefore concerns a high--low estimator for one trading pair;
its basis-point scale must not be interpreted as executable transaction costs.

More broadly, the analysis is an audit-led reconstruction rather than a confirmatory test. The
positive stratified result shows that a residual contrast can remain under one coarse,
dependence-aware design. The parametric and two-part nulls show that it is not stable enough to name
a structural ``extremity premium.'' A multi-asset design with archived intraday quotes, a versioned
sentiment construction, and an ex ante analysis plan would be needed to adjudicate mechanism and
generalisability.

\section{Conclusion}

Extreme Fear \& Greed days have wider estimated Bitcoin spreads unconditionally. Once volatility,
lagged momentum, temporal dependence and the estimator's zero floor are exposed, the gap becomes
smaller and its inference depends on specification. We therefore answer the research question
conditionally: a positive difference remains, but it is not inferentially stable. The useful result is the boundary: this dataset
supports an exploratory, method-contingent association, not a causal liquidity claim or trading
rule. The methodological contribution is to make that boundary visible rather than select the
favourable specification.

\section*{Declarations}

\noindent\textbf{Funding.} No external funding was received for this research.

\noindent\textbf{Competing interests.} The author declares no competing interests.

\noindent\textbf{Data and code availability.} The aggregate Fear \& Greed series is provided by
Alternative.me and the daily OHLCV inputs by Binance. Their cached values, reconstruction scripts,
processed panels and deterministic outputs are included in the accompanying replication package.
The provider's unpublished index components prevent reproduction of its upstream construction.
The repository is \url{https://github.com/dissensus-ai/extremity-premium}. The analysis code and inputs are available at commit
\href{https://github.com/dissensus-ai/extremity-premium/tree/56177c999d63cc7b9842a5c1c55645680fbe2c35}{\texttt{56177c9}}. The release manifest identifies the exact
code state and file hashes.

\noindent\textbf{Declaration of generative AI use.} Anthropic Claude, Google Gemini and OpenAI GPT
models assisted during earlier drafting, code review and statistical exposition. OpenAI Codex
assisted in September 2026 with code and result auditing, statistical script development, citation
checking, restructuring and language revision. The author retains responsibility for the
manuscript, its claims and the decision to submit it.

\clearpage
\section*{Correction and relationship to earlier versions}

This reconstruction supersedes \emph{The Extremity Premium},
\href{https://arxiv.org/abs/2602.07018}{arXiv:2602.07018}. The illustrative agent-based model and its moment-matching discussion have been
removed; the model's spread--uncertainty link was imposed by its rules and did not independently
identify a mechanism. The uncertainty-component decomposition (including the earlier 81.6\%/18.4\%
allocation) and associated contribution are withdrawn from the paper. The earlier Abdi--Ranaldo
regime row, coefficient $+0.048$ with $p=.003$, is withdrawn because it was contradicted by the
committed analysis; it must not be cited as replication. Other untraceable robustness/rolling-window
rows are omitted rather than treated as evidence.

The former momentum control used Parkinson volatility as its outcome and could not establish
spread robustness. The replacement uses spreads directly and reports both the parametric null and
the positive stratified contrast. Sample-size-weighted pooled standard deviations replace the
legacy unweighted effect-size denominators: the raw BTC and ETH effects are $.374$ and $.290$
(previously $.404$ and $.312$), and the close-to-close-volatility-demeaned BTC effect is $.154$
(previously $.167$). The former $.21$ headline concerned a different range-volatility/positive-
estimate specification and is not the headline reported here. Apparent quote/effective-spread
validation is also withdrawn for the provenance and construct failures explained in the supplement.
These corrections change the conclusion: this paper supports an exploratory association whose
inference depends on specification, not identified adverse selection, causal liquidity withdrawal,
or a trading recommendation. The expanded historical source is retained for audit and should not
be used as the current manuscript.

\clearpage
\renewcommand{\theHsection}{S\arabic{section}}
\renewcommand{\theHtable}{S\arabic{table}}
\renewcommand{\theHequation}{S\arabic{equation}}
\setcounter{section}{0}
\setcounter{table}{0}
\setcounter{equation}{0}
\renewcommand{\thesection}{S\arabic{section}}
\renewcommand{\thetable}{S\arabic{table}}
\renewcommand{\theequation}{S\arabic{equation}}
\begin{center}{\Large\bfseries Supplementary material}\end{center}
\section{Data provenance and sample construction}

The analysis reconstructs the daily panel from cached upstream CSVs in
\path{data_ingestion/replication_inputs/} and verifies it against
\path{results/full_sample_btc_data.csv}. Its market fields originate from Binance's public
BTC/USDT daily OHLCV endpoint; its aggregate Fear \& Greed value, provider class and timestamp
originate from Alternative.me's public endpoint. The panel contains 2,896 ordered observations
from 21 February 2018 through 29 January 2026. There are four absent calendar dates in two gaps
(14--16 April 2018 and 26 October 2024); they are not imputed. API observations are timestamped at
UTC midnight and are joined to Binance candles by UTC date. The provider does not document exact
release timing, so no no-look-ahead or prediction claim is made.

The primary classification is implemented by
\texttt{analysis/full\_sample\_extension.py}: extreme fear $\leq25$; fear 26--45; neutral 46--55;
greed 56--75; extreme greed $>75$. Counts are 607, 808, 457, 743 and 281, respectively. The two
extremes are pooled for 888 observations. These cut-offs are author choices. The provider's cached
classes supply 391 neutral observations (47--54 in this panel), while the extreme group is
unchanged. After the strictly lagged momentum warm-up, the analysis sample runs from 13 March 2018
through 29 January 2026 ($N=2{,}876$); the primary and API neutral counts are 454 and 389.

Alternative.me's current methodology page names volatility/maximum drawdown, market
momentum/volume, social media, surveys, Bitcoin dominance and Google Trends as inputs. Surveys are
marked paused. Public output contains the aggregate score, class and timestamp, not component
histories, the operational treatment of the paused component, a reconstructible aggregation
formula or version history. The study can reproduce its calculations conditional on those public
aggregate scores; it cannot reproduce the upstream index.

\section{Outcome construction and estimator floor}

The index and exchange series are joined before rolling features are computed, so the four
missing dates are also skipped by the return/rolling calculations. A return crossing either gap
spans more than one calendar day. The 20-observation windows below should therefore not be read as
exactly 20 calendar days throughout. For adjacent observed rows $t-1,t$, the code constructs
\begin{align}
\beta_t &= \left[\log(H_{t-1}/L_{t-1})\right]^2+
           \left[\log(H_t/L_t)\right]^2,\\
\gamma_t &= \left[\log\!\left(\max(H_{t-1},H_t)/
                 \min(L_{t-1},L_t)\right)\right]^2,\\
\alpha_t &= \frac{\sqrt{2\beta_t}-\sqrt{\beta_t}}{3-2\sqrt{2}}
            -\sqrt{\frac{\gamma_t}{3-2\sqrt{2}}},\\
\widehat S_t &= 10{,}000\times
  \frac{2\{\exp[\max(\alpha_t,0)]-1\}}
       {1+\exp[\max(\alpha_t,0)]}.
\end{align}
This is the implemented Corwin--Schultz high--low estimator, expressed in basis points. The
implementation follows the authors' negative-estimate floor. Of the 2,896 observations, 935
(32.3\%) are zero-valued after this floor. Within the momentum-eligible sample, 926 of 2,876
(32.2\%) are floored. These are boundary values of an estimator, not observations of a literally
zero market spread.

The 20-day realised-volatility control is the rolling standard deviation of close-to-close log
returns multiplied by $\sqrt{252}$. Momentum for observation $t$ is
\[
M_t=\sum_{j=1}^{20}r_{t-j},
\]
so the current return is excluded. Realised volatility and momentum are standardised over the
analysis sample before their squared terms are constructed.

\section{Specification ladder and inference}

The unconditional comparison and the pooled realised-volatility-quintile residual comparison use
IID Welch tests. The raw interval is a seeded percentile bootstrap interval. These legacy
comparisons are descriptive because they do not address serial dependence. The five separate
quintile comparisons use Welch intervals and Holm-adjusted $p$-values.

The quadratic model includes an intercept, separate extreme and neutral indicators, realised
volatility, realised volatility squared, lagged momentum and lagged momentum squared. Moderate fear
and greed are the common reference. The reported extreme-minus-neutral result is a linear contrast
of the two indicator coefficients. OLS spread-level and log-positive-magnitude models use a
Newey--West covariance with 20 ordered-observation lags. The binomial positive-incidence model uses
the corresponding HAC covariance. For the positive-magnitude model, zero score contributions retain
floored observations' positions in the ordered series.

The non-parametric design fixes realised-volatility quintiles crossed with momentum terciles. In
each of 15 cells it computes the extreme-minus-neutral gap; cell weights are
$n_E n_N/(n_E+n_N)$. The reported statistic is their weighted average. Its two-sided reference
distribution contains the statistic from each of the 2,875 non-zero circular shifts of the full
regime sequence, with the observed alignment included in the $p$-value denominator. This design
preserves regime runs and outcome order, conditional on the fixed covariate cells. It relies on
phase-exchangeability and can fail under slowly changing omitted states.

\begin{table}[ht]
\centering
\caption{Main specification ladder reproduced from source CSVs}
\footnotesize
\begin{tabular}{L{4.3cm}rL{3.1cm}L{2.7cm}r}
\toprule
Specification & $N$ & Effect & 95\% interval & $p$ \\
\midrule
Unconditional mean gap & 2,896 & +61.99 bps ($d=0.374$) & [46.60, 77.89] & $2.70\times 10^{-14}$ \\
RV-quintile demeaning & 2,896 & +24.79 bps ($d=0.154$) & -- & 0.0015 \\
Quadratic RV + momentum, HAC & 2,876 & +11.81 bps & [-2.31, 25.93] & 0.101 \\
Fixed 5$\times$3 stratification & 2,876 & +20.44 bps & -- & 0.0195 \\
Positive-estimate incidence, HAC & 2,876 & OR 1.045 & [0.806, 1.356] & 0.739 \\
Positive-estimate magnitude, HAC & 1,950 & +11.3\% & [-4.8\%, 30.2\%] & 0.180
\\
\bottomrule
\end{tabular}
\end{table}

\section{Individual realised-volatility quintiles}

\begin{table}[ht]
\centering
\caption{Extreme-minus-neutral gaps within close-to-close realised-volatility quintiles}
\footnotesize
\begin{tabular}{rrrrrrr}
\toprule
Quintile & $n_E$ & $n_N$ & Gap (bps) & 95\% CI & Raw $p$ & Holm $p$ \\
\midrule
Q1 & 58 & 173 & +11.78 & [-10.73, 34.28] & 0.301 & 0.540 \\
Q2 & 134 & 103 & +21.35 & [-4.95, 47.65] & 0.111 & 0.540 \\
Q3 & 159 & 65 & +24.37 & [-8.09, 56.83] & 0.140 & 0.540 \\
Q4 & 236 & 74 & +23.72 & [-11.50, 58.93] & 0.185 & 0.540 \\
Q5 & 301 & 42 & +49.35 & [-11.11, 109.80] & 0.108 & 0.540
\\
\bottomrule
\end{tabular}
\caption*{\footnotesize All intervals cross zero and 0/5 tests reject after Holm adjustment. The
pooled residual comparison in the main ladder is an exploratory aggregation, not a replacement for
these five tests.}
\end{table}

\section{Quadratic-model coefficients}

\begin{longtable}{L{3.3cm}L{4.0cm}rrr}
\caption{Full coefficients for the custom 46--55 classification}\label{tab:coefficients}\\
\toprule
Model & Term & Estimate & HAC SE & $p$ \\
\midrule
\endfirsthead
\toprule
Model & Term & Estimate & HAC SE & $p$ \\
\midrule
\endhead
Spread level & Constant & +108.4195 & 3.8216 & $4.74\times 10^{-177}$ \\
Spread level & Extreme & +8.2217 & 6.1979 & 0.185 \\
Spread level & Neutral & -3.5895 & 5.5290 & 0.516 \\
Spread level & Realised volatility & +40.3750 & 3.8579 & $1.24\times 10^{-25}$ \\
Spread level & Realised volatility squared & -4.0588 & 0.9287 & $1.24\times 10^{-5}$ \\
Spread level & Lagged 20-day momentum & -2.6958 & 3.9375 & 0.494 \\
Spread level & Lagged momentum squared & +10.2502 & 2.7357 & 0.0002 \\
Positive incidence & Constant & +0.6957 & 0.0615 & $1.10\times 10^{-29}$ \\
Positive incidence & Extreme & +0.0079 & 0.0952 & 0.934 \\
Positive incidence & Neutral & -0.0364 & 0.1113 & 0.744 \\
Positive incidence & Realised volatility & -0.1419 & 0.0567 & 0.0123 \\
Positive incidence & Realised volatility squared & +0.0296 & 0.0093 & 0.0014 \\
Positive incidence & Lagged 20-day momentum & -0.0209 & 0.0411 & 0.611 \\
Positive incidence & Lagged momentum squared & +0.0250 & 0.0277 & 0.366 \\
Log positive magnitude & Constant & +4.7556 & 0.0353 & $<10^{-300}$ \\
Log positive magnitude & Extreme & +0.0650 & 0.0582 & 0.264 \\
Log positive magnitude & Neutral & -0.0422 & 0.0604 & 0.485 \\
Log positive magnitude & Realised volatility & +0.4314 & 0.0351 & $8.40\times 10^{-35}$ \\
Log positive magnitude & Realised volatility squared & -0.0601 & 0.0093 & $1.27\times 10^{-10}$ \\
Log positive magnitude & Lagged 20-day momentum & +0.0140 & 0.0267 & 0.601 \\
Log positive magnitude & Lagged momentum squared & +0.0473 & 0.0179 & 0.0082
\\
\bottomrule
\end{longtable}

\noindent Coefficients in the logit model are log odds; coefficients in the positive-magnitude
model are log ratios. The main paper reports the extreme-minus-neutral contrasts, not either
indicator coefficient in isolation.

\clearpage
\section{Classification sensitivity}

\begin{table}[ht]
\centering
\caption{Results using the cached API-provided class labels}
\footnotesize
\begin{tabular}{L{5.8cm}rL{2.5cm}r}
\toprule
Specification & $N$ & Effect & $p$ \\
\midrule
Quadratic spread level, HAC & 2,876 & +11.36 bps & 0.119 \\
Positive-estimate incidence, HAC & 2,876 & OR 1.018 & 0.899 \\
Positive-estimate magnitude, HAC & 1,950 & +12.0\% & 0.175 \\
5$\times$3 stratified spread & 2,876 & +20.12 bps & 0.0282 \\
5$\times$3 stratified incidence & 2,876 & +1.47 pp & 0.624 \\
5$\times$3 stratified magnitude & 1,950 & +11.8\% & 0.194
\\
\bottomrule
\end{tabular}
\caption*{\footnotesize Spread effects are in basis points; incidence effects are odds ratios for
the parametric model and percentage-point differences for the stratified model; magnitude effects
are percentage differences.}
\end{table}

\section{Bin-resolution and shared-input sensitivity}

\begin{longtable}{L{4.2cm}rrrrr}
\caption{Pooled demeaned gaps across volatility axes, filters and bin counts}\label{tab:bins}\\
\toprule
Variant & Bins & $N$ & Gap (bps) & $d$ & IID $p$ \\
\midrule
\endfirsthead
\toprule
Variant & Bins & $N$ & Gap (bps) & $d$ & IID $p$ \\
\midrule
\endhead
parkinson\_filtered & 5 & 1,961 & +30.07 & 0.196 & 0.0008 \\
parkinson\_filtered & 10 & 1,961 & +20.43 & 0.134 & 0.0223 \\
parkinson\_filtered & 20 & 1,961 & +16.27 & 0.107 & 0.0668 \\
parkinson\_filtered & 50 & 1,961 & +15.90 & 0.107 & 0.0676 \\
parkinson\_filtered & 100 & 1,961 & +15.21 & 0.103 & 0.0787 \\
parkinson\_unfiltered & 5 & 2,896 & +21.28 & 0.132 & 0.0063 \\
parkinson\_unfiltered & 10 & 2,896 & +16.61 & 0.104 & 0.0327 \\
parkinson\_unfiltered & 20 & 2,896 & +14.48 & 0.091 & 0.0613 \\
parkinson\_unfiltered & 50 & 2,896 & +12.45 & 0.078 & 0.108 \\
parkinson\_unfiltered & 100 & 2,896 & +12.97 & 0.082 & 0.0923 \\
realized\_filtered & 5 & 1,961 & +34.72 & 0.223 & 0.0001 \\
realized\_filtered & 10 & 1,961 & +27.02 & 0.175 & 0.0028 \\
realized\_filtered & 20 & 1,961 & +26.87 & 0.176 & 0.0028 \\
realized\_filtered & 50 & 1,961 & +27.98 & 0.185 & 0.0017 \\
realized\_filtered & 100 & 1,961 & +27.51 & 0.184 & 0.0018
\\
\bottomrule
\end{longtable}

\noindent ``Parkinson'' is a range-based volatility axis which shares high--low prices with the
spread estimate. ``Filtered'' additionally removes zero-floored spread estimates. The table is a
sensitivity grid, not a family of independent confirmatory tests. In particular, the
Parkinson-unfiltered IID $p$-value crosses .05 between 10 and 20 bins. The realised-filtered rows
remain small, but answer a different question on a selected positive-estimate sample.

\section{Withdrawal of quote-validation claims}

The earlier short-window comparisons are not retained as independently observed quote or effective-
spread validation. Inspection of a nominal Bybit order-book file found only trade fields
(timestamp, symbol, side, size, price and trade identifiers), while the processing code permits
fallback to within-interval trade ranges. The 90-day summary has no committed dedicated regenerator.
The Binance comparator computes twice the absolute price deviation from a 100-trade rolling median,
scaled by that median and trimmed above the daily 99th percentile. It does not use an observed
quote midpoint. The historical comparisons are available in the repository for audit, but their
labels do not establish construct validity. A new validation would require archived bid and ask
quotes with a documented sampling and aggregation path.

\section{Reproduction map}

\begin{longtable}{L{5.7cm}L{8.1cm}}
\toprule
Artifact & Role \\
\midrule
\endfirsthead
\toprule
Artifact & Role \\
\midrule
\endhead
\path{analysis/spread_momentum_hurdle.py} & Repaired spread-side quadratic, floor and circular-shift analyses.\\
\path{results/spread_momentum_hurdle_parametric.csv} & Main contrasts, standard errors, intervals and HAC $p$-values.\\
\path{results/spread_momentum_hurdle_nonparametric.csv} & Fixed-grid estimates, circular-shift $p$-values and reference quantiles.\\
\path{results/spread_momentum_hurdle_coefficients.csv} & All quadratic-model coefficients.\\
\path{results/spread_momentum_hurdle_descriptives.csv} & Floor counts and group distributions.\\
\path{analysis/headline_volatility_control.py} & Regenerates the pooled quintile-demeaned comparison and corrected $d$.\\
\path{analysis/full_sample_extension.py --effect-sizes-only} & Regenerates corrected raw and subgroup effect sizes from cached panels.\\
\path{analysis/build_frl_candidate_tables.py} & Selects named results and rebuilds every table row in this package.\\
\path{results/frl_specification_ladder.csv} & Machine-readable version of the main letter table.\\
\bottomrule
\end{longtable}

\noindent The accompanying replication package includes cached upstream inputs, an environment lock,
a code-state identifier and file hashes. It reconstructs the input panel and regenerates all
retained table rows without live API calls. The analysis code and inputs are publicly available at commit \texttt{56177c9}.

\end{document}